Authentic Research in the Classroom for Teachers and Students

L. M. Rebull (Caltech-IPAC/IRSA and NITARP)

**Abstract.**

With the advent of research-grade robotic telescopes (and professional archives) coupled with the wide availability of the Internet in schools, getting high-quality data in the classroom has become much easier than ever before. Robotic telescopes (and archives) have revolutionized what is possible to accomplish in the confines of a high school classroom. Especially in the context of new science standards in the US, schools need to be moving towards more project-based learning and incorporating more authentic scientific inquiry, so demand for programs such as this is only expected to grow. This contribution highlights a few of the programs that incorporate authentic research in the classroom, via teachers and/or students. I also point out some recurring themes among these programs and suggest a funnel as a way to think about the 'ecosystem' of projects getting astronomical data into the hands of teachers, students, and the public.

**Introduction.**

In recent years, more and more schools in the US have installed high-speed internet connections and are doing more to incorporate computers and even real scientific data into the classroom (see, e.g., Fitzgerald et al. 2004 for a review). Astronomy is particularly suited to inquiry and project-based learning in the classroom because so much data are online, so many archives are publically accessible, and so many telescopes (and telescope networks) offer time to educators. With the advent of the Next Generation Science Standards (NGSS 2013; also see A Framework for K-12 Science Education, NRC 2012), demand for such programs will likely grow (at least in the US).

In the context of the RTSRE conference, the description of the goals for the conference include the words, "Remotely located... small, optical robotic telescopes... and high school and undergraduate students..." In the context of the present contribution, I have expanded this definition slightly. Many robotic telescopes are remotely-located in orbit. Many telescopes observe in wavelengths that are not optical; several contributions to the conference focused on radio data (e.g., Levin et al. or Hollow et al., this volume). Some of the existing programs work directly with students, but some work with teachers. The programs that work with teachers take advantage of the fact that if you change the way a teacher thinks about science (and scientists), you can influence all the students a classroom teacher comes into contact with this year, next year, and the rest of her career. Additionally, I emphasize that increasingly, astronomical archives include more and more data, and more and more high-quality, ready-to-use data products.

**The Funnel of Interest.**

In contexts beyond astronomy, many fields use the model of a funnel. For example, in marketing, the funnel model is widely used to describe enticing people to buy a product; Strong (1925) cites E. S. Lewis in 1898 for the original idea.

There are a wide variety of programs using astronomy data in the classroom, and I can thick of them as an ecosystem. There are roughly four categories in this ecosystem, each with different audiences, challenges, and goals:
- **Citizen Science**: Doing something small to contribute to the whole. (Everyone plays a small role; participation does not require an understanding of the bigger picture; people are still excited that they are participating.)
- **Using Real Data**: Reproductions of simple or done projects, using real data (professional or really good amateur); an example might be to rediscover Hubble's Law using a particular data set.
- **Contributing real data**: Doing a project using new data collected for the project or a combination of new and archival data. An example might be to monitor a star, add new data to prior monitoring, and find the planet that is known to be there.
- **Original research:** Doing an original research project using professional quality new or archival data.

Just as in the biological definition of ecosystem, each component of the astronomy-data-in-the-classroom ecosystem is important and valid and worthy; each has a different footprint and reaches a different audience of educators and students and the public.

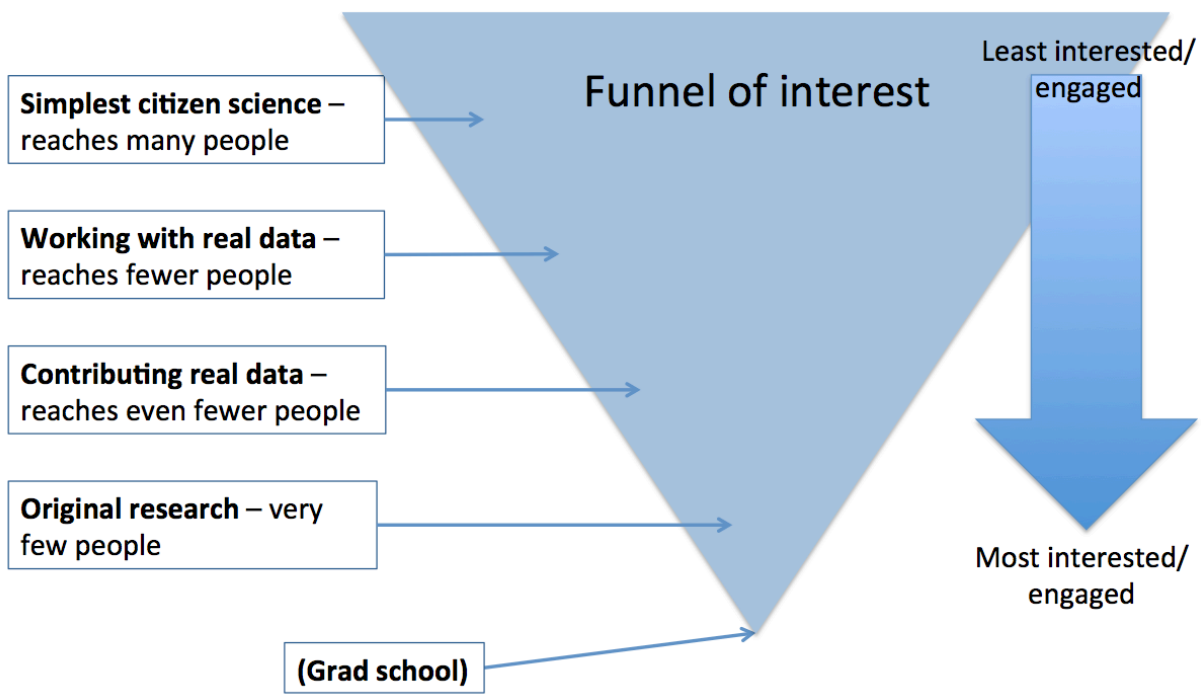

*Figure 1: The funnel of interest in the astronomy data in the classroom ecosystem. All parts of the ecosystem are important. The programs at the top reach the most people; the most interested (or engaged) people from each level of the funnel move further down the funnel.*

This ecosystem, however, can also be organized as a "funnel" of interest, from least to most interested:
- At its simplest, **citizen science** reaches many people; these people may have hints of bigger picture, but in order to participate, they don't need a deep understanding of the relevant astrophysics or of programming. Young children through senior citizens, worldwide, can participate and be excited about contributing to science.
- **Working with real data** reaches fewer people because to work with real data, participants need a deeper understanding of technical skills and/or programming, and they also need at least some understanding of the relevant astrophysics.
- **Contributing real data** reaches even fewer people; participants need to understand deeply what they are doing at least with their own data and how their data fit in to the larger astrophysical problem.
- Very few people can participate in **original research**, because in order to do research, they need a very deep understanding of what they're doing, including the how and why, as well as technical skills such as programming.

The most interested participants in citizen science programs may be interested in going deeper and working more directly with real data, i.e., move down the funnel. The most enthusiastic people working with real data may be inspired to contribute real data (using their own telescope or another accessible telescope). After contributing data, one then might be enticed to participate in original research with real data. The most motivated participants are then likely to want to come up with their own ideas for research using data they collect; the bottom of the funnel in this model, then, is grad school.

While I describe this as more or less a continuum in the context of the rest of this paper, I note for completeness that it could be argued that this is not a one-dimensional continuum, where other dimensions could include level of student involvement, or degree of instrumentation building, etc.

**Examples of projects.**

In this section, I list just a small handful of projects to demonstrate how I see the funnel as being populated. This list here is not complete, but I have attempted to create a more complete list of programs (world-wide, but just astronomy) at
http://nitarp.ipac.caltech.edu/page/other_epo_programs
I welcome all contributions and corrections.

*Citizen Science: Zooniverse and Disk Detective.*

Citizen Science at its plainest doesn't require extreme astrophysics or technical (computer) skills, so everyone can participate, and it is real science in that it ultimately can result in refereed journal articles (e.g., Schmitt et al. 2016 or Boyajian et al. 2016). It is, generally, removed from the scientific process, which is in reality messier than simplified interfaces permit. But Citizen Science at its best enables "hooks" for interested people to go further; an astronomy project might put the position (RA and Dec) of a given object on the same screen seen by all citizen scientists, which enables searching by position (in, say, SIMBAD) for those who want to learn more about that object.

Zooniverse (https://www.zooniverse.org/) is an example of this; they have many astronomy projects to choose from. Lesson plans provide a framework for canned use of the data, supporting those educators just getting started at the top of the funnel. One Zooniverse project, Disk Detective (Kuchner et al. 2016), places the RA and Dec of each object, along with a link to SIMBAD, on each of their object pages. This project, however, basically used Zooniverse to find the few most highly interested/capable amateurs worldwide to work intensively on their project. Now, they have weekly science meetings that include these most capable volunteers, involving them more intensively in the real science. This team is already helping people work their way down the funnel, but note that this is within their project and this is not necessarily something that all Zooniverse projects do. Disk Detective primarily uses archival professional (mostly IR) data, using their interface; the inner circle of dedicated volunteers participates in getting new data.

*Working with real data: SDSS Voyages*

The Sloan Digital Sky Survey (SDSS) Voyages (http://voyages.sdss.org/ ) and Sky Server projects (http://cas.sdss.org/dr5/en/proj/ ) provide activities that work with professional data from SDSS, using the same interface professionals use. These activities cover introductory through advanced materials; participants are skill building and scaffolding so that they can go farther if they want. Users can simply work through the examples that are posted on the web; motivated participants then have powerful skills to extend to more SDSS projects, or to apply elsewhere, and move down the funnel. No special software is needed, per se, though users need to find a way to keep track of (and plot) data. No interaction with the SDSS staff is required or expected; the exercises are standalone.

*Working with real data: RBSEU*

Research Based Science Education for Undergraduates  (RBSEU, http://rbseu.uaa.alaska.edu/ ) is based at the University of Alaska at Anchorage and Indiana University. This program is a descendent of the RBSE and TLRBSE programs that were aimed at educators and run by NOAO (see, e.g., description in Fitzgerald et al. 2014). Their current goal is to test the effectiveness of research as part of undergraduate courses (see, e.g., Wooten et al. in press).  They have several battle-tested exercises using real (optical) data from several different professional telescopes; some are easily extended to use new data, in which case student results using new data have the potential to be incorporated into the scientific literature.  This project uses ImageJ, Graphical Analysis, etc.

*Working with real Data: PSC*

The Pulsar Search Collaboratory (PSC; Blumer et al. this volume, also see Rosen et al. 2010 and Blumer et al. 2018) is run by West Virginia University and NRAO and GBT ( http://psrsearch.wvu.edu, http://pulsarsearchcollaboratory.com/). The original data that were used were obtained by the Green Bank Telescope, which was drift scanning during downtime; since that initial effort, more data have been added. Participants look for new pulsars in these professional radio data, and refereed journal articles have resulted (e.g., Swiggum et al. 2015). Some participants obtain follow-up data using other radio telescopes. There are no generally available materials; users have to join to get trained and get data access. The software used is PRESTO.

*Working with real data: IASC*

The International Astronomical Search Collaboration (IASC; http://iasc.hsutx.edu/ ; see Miller et al. this volume or Miller et al. 2008) is run by Hardin-Simmons University and collaborators. Their goal is to find/refine orbits for asteroids, NEOs, comets; results are submitted to the Minor Planet Center (MPC). They use optical data from many professional telescopes, organized around intensive 30-60d campaigns. They provide online training for international teachers and students; for software, they use Astrometrica.

*Working with/Contributing real data: Microobservatory*

The Microobservatory at Harvard/CfA has been operating for many years (Sienkiewicz et al. or Dussault et al., this volume; see also, e.g., Gould et al. 2006). Here, I highlight their "Laboratory for the Study of Exoplanets" (https://www.cfa.harvard.edu/smgphp/otherworlds/ExoLab/ ).
Participants use data from their archives or that they obtain via robotic (optical) telescopes from any of a set of known, bright enough exoplanet host stars. They have several well-thought-out lessons to take users through the project. They provide a transit calendar so that users are likely to succeed unless weather intervenes. Users work within the project's web-based framework.

*Contributing real data: GEONS*

GEONS is the Geomagnetic Event Observation Network by Students (http://cse.ssl.berkeley.edu/artemis/epo-geons-program.html ; see Craig et al. 2005) This project is run by the THEMIS education and public outreach team at Berkeley; their goal is studying the interaction between Solar wind and Earth. The organizers send participants magnetometer stations, but participants have to be rural and remote enough that they will get reasonable data. No special software is required.

*Contributing real data: AAVSO*

The American Association of Variable Star Observers (AAVSO; https://www.aavso.org/ ) has a long history of supporting 'amateurs' in their studies of variable stars (Percy et al., and Kafka et al., this volume; also see, e.g., Percy 2016). Participants use their own telescopes, or they can use their data from other high-quality 'amateur' telescopes. The AAVSO has many activities, many ways to learn (online, training, local mentors), and many ways for anyone to contribute to our understanding of variable stars. People who contribute data must understand the requirements for high-quality data, and do their own photometry; there is no standardized software package for this that all AAVSO participants use. The AAVSO also maintains its own refereed journal, which is unusual for any program in the funnel; a refereed journal for student/teacher work from the broader community is a clear need.

It is perhaps worth noting that the AAVSO has crossed the boundary of sharing substantial data with the professional community. The AAVSO Photometric All-Sky Survey (APASS; Henden & Munari 2014) has gained the trust of the professional community as a reliable data source and data from it are appearing in refereed journal articles more and more frequently. No other education project of which I am aware contributes data to the professional community, certainly not on this scale.

*Original research: NITARP*

NITARP is the NASA/IPAC Teacher Archive Research Program (http://nitarp.ipac.caltech.edu ; Rebull et al. this volume), run from Caltech/IPAC. NITARP partners small groups of educators with a research astronomer for a year-long authentic research project. All teams must use data housed at IPAC (mostly infrared data, mostly NASA, all professional data). This program is aimed at teachers, but teachers may involve their students in the project at their discretion. NITARP culminates in going to an American Astronomical Society (AAS) meeting with a science poster (that appears in a science session), so participants are treated like all other professional astronomers there. All teams present posters, but, as for professional astronomers, not every poster results in a refereed journal article; some NITARP journal articles include Rebull et al. (2015, 2013, 2011). Teams typically use software that is broadly available like Excel; the Aperture Photometry Tool (APT; Laher et al. 2012ab) was developed several years ago, and is still used by NITARP teams.

*Original research: RETs*

RETs are Research Experiences for Teachers; RETs are the educator analog to the very popular REUs (Research Experiences for Undergraduates). These are all NSF funded. They no longer seem to be broadly available in astronomy, though they are flourishing in other fields. The organizing concept is that the RET provides not only the research experience, but also some support structures; it is a summer experience, paid, at a site with many RETs, and there are organized events during the summer to unify the cohort and provide support. These programs are solely focused on teachers; no students are explicitly involved per se. In RETs, software and data used vary.

**Recurring themes.**

In this section, I identify some recurring themes that are found across all of these projects – all projects known to me, not just the ones listed above as examples.

*Goals*

Everyone has broadly similar goals. These projects are trying to give people a better sense of how science really works, letting people participate in science at some level or even "peek behind the curtain." As a secondary goal, many cite the development of critical thinking skills, computer skills, and/or, engineering skills. Many of these programs provide some linkages to national standards, but at the same time they are not strictly constrained by them. An implication of this, however, is that teachers at schools that are strictly limited by standards may not be able to (may not be allowed enough flexibility to) participate in these programs.

However, note that while these programs have broadly similar goals, they often have very different specific goals. This becomes obvious when considering how different the target audiences are for these programs. A citizen science project (at the widest part of the funnel) may aim to engage people numbering in the thousands to hundreds of thousands per year; a program enabling authentic research (near the bottom of the funnel) may only work with less than a dozen new educators per year. The ratio of organizer time per participant may be low near the top of the funnel, and very high near the bottom of the funnel. Some of these programs are aimed at anyone with sufficient interest to keep coming back to a web page for a few minutes at a time; some require a year (or more) of intense engagement. Some are aimed at teachers, or teachers with students, or just students. Some may require work in teams; some may be more effective with individuals. Those programs that have a higher ratio of organizer time per participant also require more out of their participants; I suspect such programs may also make a bigger impact on their lives. (For example, a NITARP teacher told me, "I lay awake at night thinking about data"; see Rebull et al. 2018.)

*Project management*

Essentially all of these projects have difficulties finding funding for operations, or for formal evaluation (see, e.g., Buxner et al., this volume) or formal education research. At least, projects should learn from their participants and refine the program as necessary. Finding time, funding, and an appropriate location in which to share their lessons learned is also a challenge.

The best of these projects have teams that include both scientists and educators. Without a well-integrated team, or at least a mechanism for change informed by frequent feedback, projects run the risk of assuming what other constituencies need, want, will do with the project, or can contribute to the project. If the program successfully involves participants from group x, representatives from group x are usually involved in the management/development team. This then becomes a recommendation for programs wanting to expand and target a new group: get a representative from that group on the team.

Software has been an enormous barrier in the past, because it has often difficult to have schools install software, in part because they are largely using virus-prone Windows machines. However, online collaboration tools (such as Google Drive) are recently rapidly improving. Web-based astronomy-specific tools are likely the longer-term answer. Many archives have phenomenal web-based tools, but there are still some astronomy-specific critical capabilities missing; for example, research-grade photometry is not yet possible in a web-based tool to my knowledge. There are a few active efforts to do this, and to develop web-based tools that lower the barrier to, e.g., visually impaired students (IDATA with Skynet; see e.g., Gartner et al. 2017).

*Structure and materials*

Most of the programs are structured; participants are not just given the keys to a telescope and left to wander, meaning that this is not true 'Open Inquiry' but 'Guided Inquiry' (see, e.g., Kirschner et al. 2006). Participants largely work within the project's framework.

Many programs have developed materials that are online, with varying degrees of depth, completeness, testing, and robustness. Some projects explicitly put a "firewall" between the general public and the data/activities in that users are required to go through training before getting access to the data or lessons. Others simply post the activities online for any interested party to find and work through. Some projects are designed around intensive interaction with professional astronomers (high ratio of organizer time per participant); some are designed to have almost no interaction with the professional astronomers or anyone at the project.

Few of these projects (justifiably) are for the true novice; really only the widest part of the funnel is aimed at that community. For the rest of these projects, participants have to work within the project's structure to get up to speed and then do the tasks that are part of the project. Conversely, however, if participants want to go further, they have to be willing to walk outside the structure; they must have the confidence to do so.

Many of these projects cite the need to build confidence in their participants. It's not enough to just teach them basic skills; they need the emotional support as well. This comes up most frequently in the context of working with teachers; teachers need to develop the confidence to handle the unknown, to not know everything before working with data/students, and to know how to take on a difficult task (see, e.g., Rebull et al. 2018). One suspects that this might also be an issue for any adult learner, but the stakes are higher when one has to learn material in real time in front of (or with) students.

**Feeding the Funnel.**

Knowledge of other programs in the ecosystem can help broaden the entire community of trained teachers and students. Any program in the community (at any level in the funnel) can (and should) help advertise any other program.

An ongoing challenge for some of these projects is how to keep participants engaged after the program. Especially if there are limited resources, projects must determine what fraction of those resources goes to 'repeat customers' as opposed to new participants.

Being able to refer those repeat customers to other projects in the ecosystem (either at the same level of the funnel, or further down) allows those people to grow, and allows the original program to continue to focus on new participants.

It is unlikely that one project alone can successfully completely populate levels along the whole funnel, because the projects at different levels of the funnel require vastly different approaches and time investments per participant, and because funding is already limited. Projects that are doing well in their level or niche in the funnel should most likely continue in that niche, specifically because they have evidently established best practices for their specific goals and audience. But, projects should be aware of where they are in the funnel, as well as where other projects fall, and refer people as appropriate. Programs that take applications, for example, upon encountering applicants who are not accepted (say, because they are under-qualified, over-qualified, or not in the right location for geographically-restricted programs) can refer the applicants to specific programs at appropriate levels in the funnel.

In the future, projects now running and yet to be developed need to think about how to engage under-represented minority and differently-abled audiences. This will need to be pursued at many different levels (software, management) to be successful. Already successful programs should consider expanding audiences (make the funnel bigger overall) rather than trying to reach additional levels in the funnel.

Some projects have standalone lessons on the web that can be accessed by anyone worldwide. Simply posting the materials on the website associated with the program is very useful to students or others who find it while seeking out opportunities on their own. However, these materials could provide advertising for the program and potentially can expand the audience. One possibility is to submit standalone lessons to central repositories like astroEDU, and NASA Wavelength, with links back to the original program. However, in order to be an effective advertisement, submitted materials must work independently of the project. The more interaction a project requires (more organizer time per participant), or the more integrated/specific the materials are for the project, the more difficult it is to distribute the material widely independently of the project.

The level of interaction required is also folded into scalability of the project. For programs with high levels of time investment per participant, the only practical solution to this scalability issue is to follow the "train the trainer" model (such as the mentor networks used by the AAVSO). In a truly utopian view of the world, programs could work together to specifically integrate resources so that we can continue to draw the most interested people down the funnel. Project Y could work with the management of project Z because they really need applicants to have skill x; project Z could then work towards building skill x among their participants, with an eye towards having the most enthusiastic participants from Z later become participants in Y.

**Integration with Professional Archives.**

Professional astronomy archives are getting better and better every year. Many of the projects at all but the highest level of the funnel have teachers and students accessing the same archives with the same software or interface that the professionals use. Sometimes, there is an interface for educational purposes that is slimmed down; the software might be interacting with the professional database, but it only shows a simplified interface for educators. It was noted at more than one presentation at RTSRE (e.g., Hollow et al., this volume) that students and teachers appreciate being given access to the real interface, even if the learning curve to use it is steep. Anything simpler seems "sugar coated" and possibly "less real" to participants.

Astronomy archives (at least those at NASA, with which I am most familiar; see, e.g., Rebull et al. 2016) are currently working towards these goals: (1) having more and better tools that allow users to explore the data and make connections between data sets without having to write all of their own code; (2) integration of archives, such that once users master one archive's interface, they can use that same interface to seamlessly integrate data from other archives from other locations, other missions, and other countries; (3) and development of sharable work spaces located at the archives themselves. Good progress on achieving these goals has already been made.

These goals represent a fundamental shift from "give me the data and let me take it home to analyze" (which is the way it has worked for decades, even before there were archives at all) to "do some or all of the analysis at the archive itself and give me only a subset of the data – or even just plots – to take home" (e.g., Rebull et al. 2016). Once tools exist to do the work at the archive itself, there is much less of a need for users (professional astronomers or the education community) to write absolutely all their own code to do research. For example, several archives are experimenting with Jupyter notebooks; this enables even those with minimal coding experience to write code and interact with the archives, without having to load software on their own local machines.

All three of these goals are designed to make finding, sharing, analyzing, and using professional data easy for the scientific community, the primary audience for these archives. However, since that scientific community necessarily includes people with a broad range of backgrounds (emeritus professors to summer students), these developments also directly benefit education communities. Tools that help professionals analyze data at the archive can help *anyone* analyze data at the archive.

**Summary.**

The ecosystem of programs getting real astronomy data into the hands of students/teachers is vast, indeed. The distribution of programs can be thought of as a funnel, where the programs with the lowest entry barriers scoop up the largest number of people. All parts of the funnel are important! In general, programs that require more out of their participants reach fewer people.

All of these programs that bring astronomy data to students and teachers are trying to help the wider world get involved in science, understand how science works, develop critical thinking skills, etc. However, these programs have different specific goals, which can be seen in the total number of participants, the total time the participant spends with the program, and the ratio of organizer time per participant. Most of these programs are structured, but the best give users skills and/or resources to continue to work in a less-structured context. Everyone needs money, and everyone needs better evaluation. The best projects have management teams that include representatives of their target audience or at least a mechanism for continual feedback on and changes to the program. Especially when working with teachers, projects must build confidence in their participants.

The funnel as an organization scheme for the ecosystem can provide guidelines for how to "feed the funnel." Programs should support other programs by referring participants when appropriate. Programs should probably not try to populate all levels of the funnel themselves, but should work to expand the funnel as a whole by reaching other communities (e.g., URM). Ideally, programs should work together to help participants move down the funnel.

Professional astronomy archives are moving towards features that will directly benefit the astronomy education community.

**Acknowledgements**

Thank you to Michael Fitzgerald and Wendi Laurence for helpful comments on the manuscript.